\documentclass[conference, a4paper]{IEEEtran}

\renewcommand\IEEEkeywordsname{Keywords}

\IEEEoverridecommandlockouts

\usepackage[utf8]{inputenc}
\usepackage[T1]{fontenc}
\usepackage{textcomp}
\usepackage[
            backref=false,
            backend=biber,
			sorting = none,
			url = false,
			doi = false,
			isbn = false,
			bibstyle=ieee, %
			citestyle=numeric-comp, %
			defernumbers=false
			]{biblatex}

\usepackage[table,xcdraw,dvipsnames]{xcolor}
\renewcommand*{\bibfont}{\footnotesize} %
\definecolor{ZSWDarkBlue}{RGB}{0,89,170}
\usepackage[unicode=true,
            bookmarks=false,
            bookmarksnumbered=false,
            bookmarksopen=false,
            breaklinks=false,
            hidelinks,     %
            colorlinks=False]
            {hyperref}
\hypersetup{pdftitle={Storage Placement and Sizing in a Distribution Grid with high PV-Generation},
            pdfauthor={B. Matthiss, A. Momenifarahani, Jann Binder},
            citecolor = ZSWDarkBlue,
            linkcolor = ZSWDarkBlue}

\newbibmacro{string+doiurlisbn}[1]{%
  \iffieldundef{doi}{%
    \iffieldundef{url}{%
      \iffieldundef{isbn}{%
        \iffieldundef{issn}{%
          #1%
        }{%
          \href{http://books.google.com/books?vid=ISSN\thefield{issn}}{#1}%
        }%
      }{%
        \href{http://books.google.com/books?vid=ISBN\thefield{isbn}}{#1}%
      }%
    }{%
      \href{\thefield{url}}{#1}%
    }%
  }{%
    \href{http://dx.doi.org/\thefield{doi}}{#1}%
  }%
}
\DeclareFieldFormat{title}{\usebibmacro{string+doiurlisbn}{\mkbibemph{#1}}}
\DeclareFieldFormat[article,incollection]{title}%
    {\usebibmacro{string+doiurlisbn}{\mkbibquote{#1}}}

\bibliography{literature}        %
\usepackage{siunitx}
\usepackage{eurosym}
\DeclareSIUnit \va { VA } %
\DeclareSIUnit \var { var }
\DeclareSIUnit \eur {\text{\euro}}
\DeclareSIUnit \ekwh {\eur\per{\kilo\watt\hour}}
\DeclareSIUnit \ekw {\eur\per{\kilo\watt}}
\DeclareSIUnit \ekm {\eur\per{\kilo\meter}}
\DeclareSIUnit \kva {{\kilo\volt\ampere}}
\DeclareSIUnit\year{yr}
\DeclareSIUnit \kw{{\kilo\watt}}
\DeclareSIUnit \kwh{{\kilo\watt\hour}}
\DeclareSIUnit \batt {batt}

\usepackage{mathtools} %
\usepackage{amsmath}
\usepackage{bm}
\usepackage{amssymb}

\newcommand{\vect}[1]{\boldsymbol{\mathbf{#1}}}
\DeclareMathOperator{\var}{Var}

\usepackage[short]{optidef}

\ifCLASSINFOpdf
  \usepackage[pdftex]{graphicx}
  \DeclareGraphicsExtensions{.pdf,.jpeg,.png}
\else
\fi
\usepackage{balance}
\usepackage{todonotes}

\usepackage{ifthen}
\makeatletter
\newcounter{IEEE@bibentries}
\renewcommand\IEEEtriggeratref[1]{%
  \renewbibmacro{finentry}{%
    \stepcounter{IEEE@bibentries}%
    \ifthenelse{\equal{\value{IEEE@bibentries}}{#1}}
    {\finentry\@IEEEtriggercmd}
    {\finentry}%
  }%
}
\makeatother
\usepackage{eurosym}
\usepackage{tabularx}
\usepackage{multirow}
\usepackage{booktabs}
\usepackage{makecell}

\renewcommand\cellgape{\Gape[4pt]}

\newcolumntype{L}{>{\RaggedRight\arraybackslash}X}
\newcolumntype{C}{>{\Centering\arraybackslash}X}
\newcommand\mc[1]{\multicolumn{2}{c}{#1}}
\newcommand\mr[1]{\multirow{2}{*}{#1}}

\usepackage{tikz} 
\usetikzlibrary{calc, angles, quotes}
\usepackage[europeanresistors,americaninductors]{circuitikz}
\usepackage{kbordermatrix}

\usepackage{algorithm}
\usepackage{algpseudocode}

\usepackage{optidef}

\ifCLASSOPTIONcompsoc
  \usepackage[caption=false,font=normalsize,labelfont=sf,textfont=sf]{subfig}
\else
  \usepackage[caption=false,font=footnotesize]{subfig}
\fi

\usepackage{stfloats}
\usepackage{url}
\begin{document}
\title{Storage Placement and Sizing in a Distribution Grid with high PV-Generation}

\author{\IEEEauthorblockN{
Benjamin Matthiss,
Arghavan Momenifarahani,
Jann Binder
}
\IEEEauthorblockA{
Zentrum für Sonnenenergie- und Wasserstoff-Forschung Baden-Württemberg\\
Stuttgart, Germany \\
Email: benjamin.matthiss@zsw-bw.de}

}

\maketitle
\IEEEpubidadjcol
\begin{abstract}
With the increasing penetration of renewable resources in the distribution grid, the demand for alternatives to grid reinforcement measures rises.
One possible solution is the use of battery systems to balance the power flow at crucial locations in the grid.
Hereby the optimal location and size of the system has to be determined in regards of investment and grid stabilizing effect. In this paper the optimal placement and sizing of battery storage systems for grid stabilization in a small distribution grid in southern Germany with high PV- penetration is investigated
and compared to a grid heuristical reinforcement strategy. 
\end{abstract}


\section{Introduction}
Nowadays, the rapid increment penetration of renewable energy resources into the distribution network can cause increased stress to the distribution system such as overvoltage situations or the exceeding of line ratings. Furthermore, the sustained need to establish a balance between energy demand and supply requires new solutions to enhance the reliable operation of the power system.
Battery energy storage systems (BESS) are being proposed as a measure to enable the integration of more renewable energy generation into the distribution grid without the need for curtailing renewable generations or reinforcing the grid. 
The availability of storage also allows to maximize the energy efficiency, decrease the network losses as well as being able to deliver control or reserve energy to the grid.
In this paper, the optimal placement and sizing of a  battery energy storage system (BESS) for grid relief in a PV rich distribution grid are investigated. The method used is based on a linearized load flow method and will be tested with data from a real distribution grid.

\section{Literature Review}

In the above mentioned context, battery storage systems connected to the distribution grid are the focus of numerous studies \cite{dunn_electrical_2011,divya_battery_2009,beaudin_energy_2010,Dunn2011ElectricalES,ferreira_characterisation_2013,mateo_costbenefit_2016,}.
Batteries can be used to reduce distribution system losses, removing existing hindrances to the integration of renewable distributed generation, contributing to voltage and frequency regulation \cite{oudalov_optimizing_2007,aditya_battery_2001,}, facilitating peak shaving or decreasing the need of network expansion \cite{lazzeroni_optimal_2019}.
However, on account of the specific application and operational strategy, there is a need for an appropriate procedure to size and site the mentioned systems to minimize the costs and losses \cite{grover-silva_optimal_2018,}.
Additionally, the objective function in different methodologies implies considerable variation in outcomes of the cost-optimized placement and sizing of the BESS.
Network structure, the renewable resource positions, and line-flow limits can also have impacts on optimal storage placement \cite{bose_optimal_2012,}.
The mathematical techniques to calculate the problem of optimal siting and sizing the storage, which is in general non-convex and high dimensional, are classified into analytic techniques, artificial intelligence techniques, classical techniques or other assorted techniques \cite{grover-silva_optimal_2018,prakash_optimal_2016,}.
Motalleb \textit{et al.} \cite{motalleb_optimal_2016,}, proposed a heuristic method to find the optimal locations and capacity of multi-purpose battery energy storage system (BESS) taking account of distribution and transmission parts.
Fossati \textit{et al.} \cite{fossati_method_2015,} found the optimal power capacity of a BESS that minimizes the operating cost of the microgrid based on genetic methods.
A fuzzy expert system determines the power delivered or taken from the storage system. 
The advantage of the proposed algorithm is its easy adaptation to different types of microgrids.
In \cite{chen_optimal_2011,}, the economic optimal allocation of the energy storage is explored based on net present value using matrix-real-coded genetic algorithm techniques. The upside of this method is taking complete and overall design aspects into consideration.
In \cite{arabali_cost_2013,ahmadian_optimal_2016,} the articles give a dissertation on how to minimize the sum of operation cost and to realize the optimal site and size the BESS based on particle swarm optimization (PSO).
The objective in \cite{kerdphol_battery_2014,} is to enhance frequency control and reduce an operating cost by integrating the load shedding scheme with optimal sizing of BESS.
The results depict the better performance of frequency control based on PSO in comparison with an analytic algorithm with load shedding scheme.
However, the issue identified in heuristic methods is that they require huge calculations, and it is uncertain if they converge to the global optimal solution.
In \cite{grover-silva_optimal_2018,} a second order cone program (SOCP) convex relaxation of the power flow equations for optimal sizing and siting of BESS with a lower computational burden is presented. Hereby the objective function is formulated in two different manners: minimizing investment vs. power losses and minimizing investment vs. operational cost benefits in a variable price market.
As discussed in \cite{harsha_optimal_2015,} the load and generation balancing of interconnected renewable resources and energy storage can be controlled using a dynamic energy price. The problem is formulated in a stochastic dynamic program over a finite horizon with the aim to minimize the long-term average cost of used electricity as well as investment in storage.

\section{Cost Analysis}

In this section follows an overview of the costs considered for battery placement and grid reinforcement.
These are important to asses the optimal placement and sizing of battery systems correctly.

\subsection{Levelized Cost of Electricity}
Looking at energy costs, the concept of levelized cost of energy (LCOE) is the center of attention as a significant practical as well as a transparent method for a cost and efficiency analysis which assists to establish a comparison with respect to costs between each individual alternative for energy generation.
The key concept of LCOE, in a simplified manner, is the division of complete lifetime costs consisting of investments, operations, fuel outlay, local and financing conditions over electricity generation. 
A progressive research is conducted to elaborate the substantial attributes of LCOE, for instance, Fraunhofer ISE in \cite{kost_levelized_2018,} gives a dissertation to the present and future market evolution of clean resources such as PV, wind turbines and bio-gas and predicts the regarded LCOE for different resources till 2035 based on the scenarios defined for market expansion in Germany. The authors in \cite{shea_applied_2019} carried out research on Mauritius island and put the generation resources potential and island particular costs under consideration to find out the LCOE of technologies. The LCOE results from different resources are put into application to prioritize the energy systems in a cost-effective way. In \cite{allouhi_energetic_2019,}, the article addresses three different types of PV-systems, and LCOE enables to obtain a comparison of these types under Morocco's climate and helps the decision-making process.

\subsection{Levelized Cost of Energy Storage}The cost of energy storage plays an important role in economic decision making. 
In practice, determining a criterion to compute the storage cost is not easy. Further, in contrast with energy generation resources, energy storage technologies are being employed for different services such as a buffer in a dispatchable generation, peak shaving or to increase self-consumption. 
Therefore, the parameters which intervene in each case are different.
An easy cost comparison of storage technologies is only possible when they are considered in a common location, application, and type. 
Battery storage has the property of being deployed in a distributed fashion.
To determine the investment cost of a battery storage system, levelized cost of energy storage (LCOES) is measured in euro per charged \si{\kwh} and depends on specific characteristics such as cycle lifespan, depth of discharge and round trip storage efficiency. The LCOES provides the customer a great insight into the price per kWh and helps to choose the appropriate battery by taking economic aspects into account.
 
The computation of the LCOES follows the definition of the LCOE formulation.  For instance, in \cite{pawel_cost_2014, BelderbosAndreas2016CtLC,}, 
LCOES is formulated based on LCOE while instead of using fuel cost, 
charging cost is utilized and generated electricity is replaced by discharged electricity.

To estimate the total costs of energy storage placement correctly it is necessary to split up the costs into the power dependent and the capacity dependent costs.
In \cite{mullerEvaluationGridlevelAdaptability2017} the relative component costs of battery storage systems are examined in a long-term storage market analysis, performed since 2014.
Hereby, the total costs are split up into the cell costs, the power electronics, and the peripheral system costs.
For residential systems, the relative costs for the power electronic components result in \SI{42}{\percent} of the total system costs. 
In this paper, it is assumed that the power electronics cost rise linearly with the rated system power of the battery. 
The estimated LCOES of the battery system were taken from \cite{bansalBuildingBankableSolar} for the year 2019.
The cost used for the battery placement are shown in Table\,\ref{tab:cost_batt}.

\begin{table}[h]
	\centering
 	\caption{Battery Installation Costs For 2019.}
    \begin{tabular}{lcS[table-format=6.0, table-space-text-post=\,\si{\ekwh}]} 
 \toprule
\textbf{type} & \textbf{perc.}     & \textbf{costs/value}       \\ \midrule
capacity & \SI{42}{\percent}            & 130\,\si{\ekwh}  \\
periphery &\SI{28}{\percent}        & 87\,\si{\ekwh}  \\
power electronics & \SI{30}{\percent}  & 93\,\si{\ekw}   \\[5pt]
installation   &  -  & 20000\,\si{\eur\per\batt} \\
batt. lifetime &  -  & 10\,yr
\end{tabular}
	\label{tab:cost_batt}
\end{table}

\begin{figure*}
    \centering
    \includegraphics[width=0.9\textwidth]{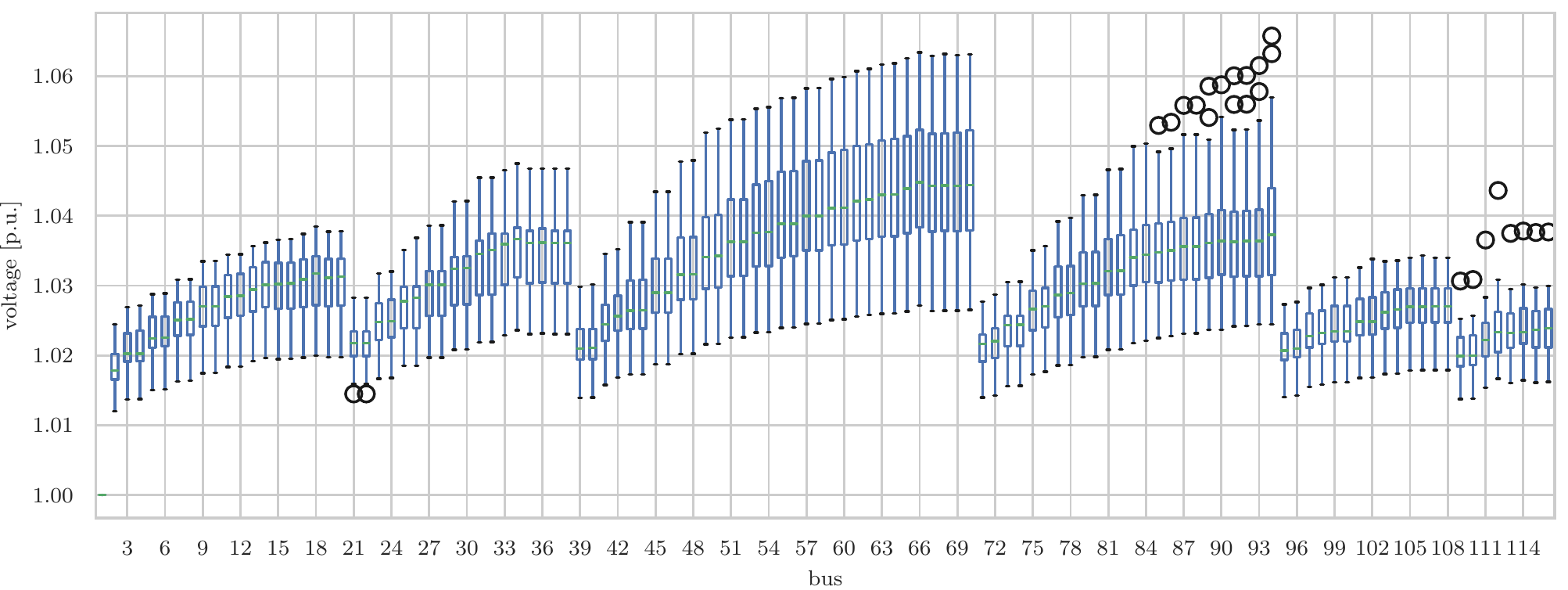}
    \caption{Voltage distribution for a day with high solar PV generation and no batter or energy curtailment. The voltage limit is 1.05 p.u., which is clearly passed }
    \label{fig:res0batt}
\end{figure*}

\begin{figure}
    \centering
    \includegraphics[width=0.35\textwidth]{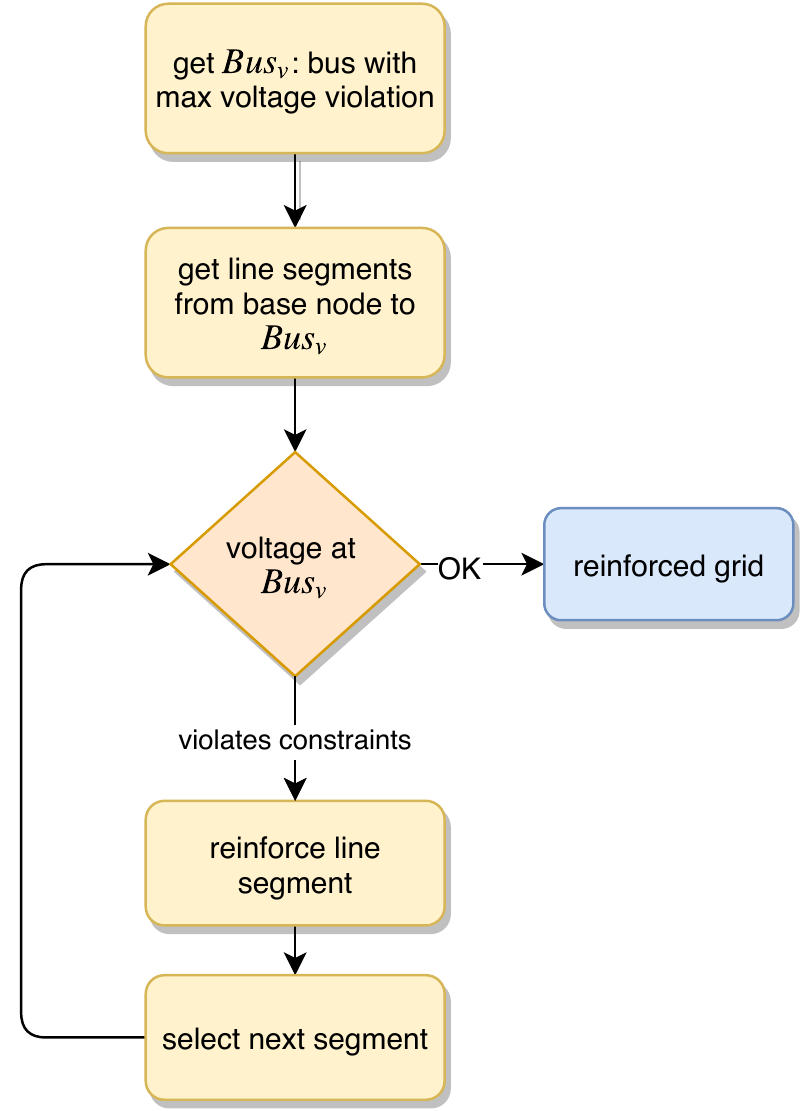}
    \caption{Flowchart of the heuristic grid reinforcement algorithm. The method shown above is repeated for all branches.}
    \label{fig:grid_re_alg}
\end{figure}

\subsection{Grid Reinforcement Costs}

For the specific expansion costs, the two main shares for cables and
installation costs are taken into account. The specific costs for the cables or lines depending on the line type are displayed in Table\,\ref{tab:cost_grid}.

However, the amounts for laying the cables vary greatly, depending on the condition of the ground. For arable land about \SI{20000}{\ekm} can be expected, but for stony ground the amount doubles (\SI{40000}{\ekm}). 
In urban areas, where roads have to be rebuilt, the costs raise up to \SI{80000}{\ekm} \cite{ruppEinflussDezentralerWaermepumpen2015}.
For the installation of another cable in parallel, it is assumed that the installation costs increase by about \SI{15}{\percent} with each added line. 

For the calculation of the annual costs, the useful lifespan of the lines still has to be determined.
For underground cables, a lifespan of 40 years are generally assumed \cite{hofmanVergleichErdkabelFreileitung2010} and the normal operating life in accordance with the Electricity Grid Charges Ordinance is also 40 years \cite{heuckElektrischeEnergieversorgungErzeugung2010}.

\begin{table}[h]
	\centering
 	\caption{Grid reinforcement costs.}
    \begin{tabular}{llS[table-format=6.0, table-space-text-post=\,\si{\ekwh}]}
\toprule
\thead{line type}                       & \thead{cost type}         & {\thead{costs}}        \\ \midrule
\multirowcell{2}{0.4 kV, $4 \times 50$ mm}  & installation & 60000\,\si{\ekm} \\
                                                        & acquisition  & 3500\,\si{\ekm} \\[5pt]
\multirowcell{2}{0.4 kV, $4 \times 120$ mm} & installation & 60000\,\si{\ekm} \\
                                                        & acquisition  & 9900\,\si{\ekm}\\[5pt]
\multirowcell{2}{0.4 kV, $4 \times 150$ mm} & installation & 60000\,\si{\ekm} \\
                                                        & acquisition  & 12000\,\si{\ekm}\\
parallel line installation                              & installation & {\makecell{additional 15\% \\ of installation costs} }  \\
Trafo, 630 kVA                                          & total        & 21000\,\si{\eur} 
\\ \bottomrule
\end{tabular}

	\label{tab:cost_grid}
\end{table}

\section{Input Data and Scenario}

In this paper real measurement data (load and irradiation) from a German distribution grid with high PV penetration is used to investigate the effects of different storage placements and grid reinforcement scenarios.

\subsection{PV Generation}
To establish a worst case scenario for the grid loading induced by distributed generation a maximum PV penetration scenario was created.
Hereby the roof area of the buildings as well as the azimuth and tilt angles were estimated to calculate the in-plane irradiance.
It was assumed that the usable roof area for PV installations is 80\,\% share of the total roof area. With this information and the knowledge about the module size and type an estimate of the maximum possible PV-capacity was established which was called $P_\text{PV, max}$.
To model the PV generation within the network the Python library \textsc{pvlib} \cite{holmgrenPvlibPythonPython2018} was used.
This package includes all necessary functions to model the complete chain from the measured irradiance (GHI, DNI, DHI) to the inverter AC output power.

\subsection{Simulation Scenario}
Because of the computational complexity of the problem (distribution grid with 106 nodes), it is not possible to optimize in a single shot over one year. 
To capture the major grid stress for the battery placement, a worst case period has been selected from the measured data. 
The selected time-span is 3 days long and shows the highest consecutive net load (generation minus load) of the whole dataset.

Different simulation scenarios have been created to evaluate the dependency on voltage limitation and PV-penetration. The selected PV penetrations are \SI{50}{\percent} and \SI{80}{\percent} relative to $P_\text{PV, max}$. The maximum allowed voltage deviation are \SI{3}{\percent} and \SI{5}{\percent} from the nominal voltage.

\begin{figure*}
    \centering
    \includegraphics[width=0.9\textwidth]{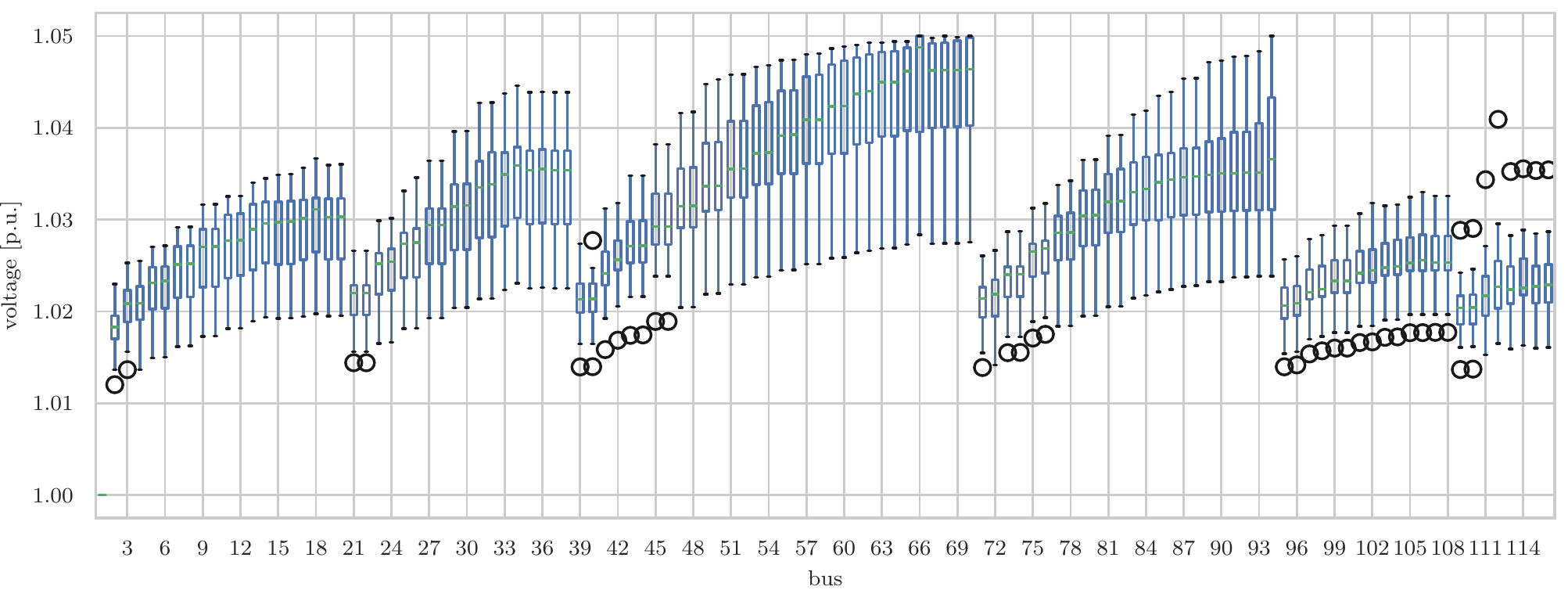}
    \caption{Voltage distribution for a day with high solar PV generation and one installed battery. In this case the voltage limit is obeyed at all nodes. The algorithm selected a battery placement at bus 65 and 93 to avoid voltage limit violations.}
    \label{fig:res1batt}
\end{figure*}

\section{Approach}

In this section we will briefly describe the methods we used for the automated grid reinforcement and the battery placement algorithm. As the specific goal of the paper is to compare optimal battery placement with grid reinforcement alternatives such as curtailment of renewable generation  to relax the grid stress are not taken into account.

\subsection{Grid Reinforcement}

The grid reinforcement was calculated with a heuristic approach.
Hereby the critical nodes of each branch are identified and the grid is reinforced until the given voltage limits are met at the corresponding nodes. The procedure of the algorithm is shown in Fig\,\ref{fig:grid_re_alg}. The algorithm chooses always the cheapest option for the reinforcement based on the costs shown in Table\,\ref{tab:cost_grid}. Hereby it is also evaluated whether it is cheaper to install multiple smaller lines in parallel or to install a larger line.

\subsection{Battery Sizing and Placement}

The algorithm is based on a linearized loadflow method presented in \cite{matthissFastProbabilisticLoad2018} and the optimization framework which is used in \cite{matthissInfluenceDemandGeneration2018b}.
The time resolution of the algorithm is 1 hour. This can be 
justified with the properties of the battery system: 
The battery capacity is larger than the 1h times the max. rated power of the battery. Furthermore, the reaction time of the battery is very fast. Therefore the battery can compensate for the fluctuations within this hour and the average power over this time span can be used for the optimization.
The simulation period as mentioned before is three days. 
Therefore multiple charging cycles are captured. This implies that the possibility to discharge the battery before the next cycle is ensured by the algorithm. Otherwise, the optimal size of the battery can not be calculated correctly.

The optimization places the battery storage units based on the costs shown in Table\,\ref{tab:cost_batt} and at the same time calculates an optimized charge trajectory for the batteries. Furthermore, constraints can be included, e.g. the number of batteries the algorithm is allowed to place.

This can be shown by an example calculation:
In Fig.\,\ref{fig:res0batt} the voltage situation within the grid is shown for a day with high solar PV generation and without an installed battery nor curtailment of PV-generation.
The voltage limit within the grid is 1.05 p.u., which is surpassed in two branches here.
In Fig.\,\ref{fig:res1batt} the result of the optimization allowing one battery and curtailment is shown: The voltage limits are obeyed at all times. The algorithm chose to place the battery at bus 65 with a size of 83 kWh. The size is determined with the help of the provided irradiation and load profiles.

\section{Results}

In this section, the results of the above-mentioned algorithms in the given simulation scenario are shown and evaluated.

\subsection{Grid Reinforcement}

\begin{figure}
    \centering
    \includegraphics[width=1\linewidth]{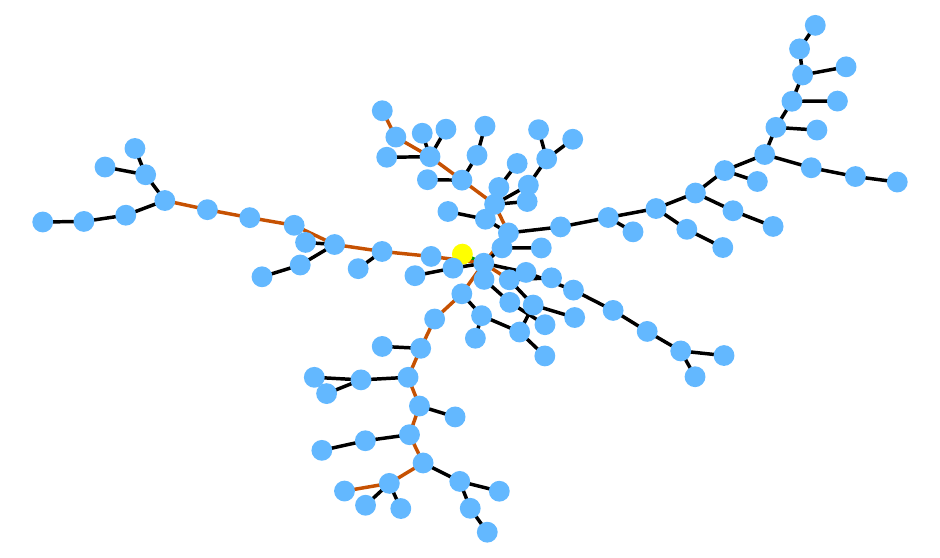}
    \vspace*{-5mm}
    \caption{Graph diagram of the reinforced grid. The yellow node indicates the transformer and all the blue nodes are busses. The red edges indicate reinforced line segments.}
    \label{fig:re_grid}
\end{figure}

\begin{table}
\sisetup{round-mode=places, round-precision=2}
\caption{Example Result of the Grid Reinforcement Algorithm: Overview of the installed assets and costs}
\label{tab:line_re}
\centering

\begin{tabular}{rrclr}
\toprule
 from bus &        to bus &  n parallel &           type &          cost [k€] \\
\midrule
  \num{1} & \num{105} & \num{2} &  NAYY $4\times120$ SE & \num{109.5} \\
  \num{1} &  \num{73} & \num{2} &  NAYY $4\times120$ SE & \num{109.5} \\
  \num{1} & \num{106} & \num{3} &  NAYY $4\times120$ SE & \num{131.1} \\
  \num{2} &  \num{28} & \num{1} &  NAYY $4\times120$ SE &  \num{87.9} \\
  \num{2} & \num{104} & \num{2} &  NAYY $4\times120$ SE & \num{109.5} \\
  \num{3} &  \num{61} & \num{1} &  NAYY $4\times120$ SE &  \num{87.9} \\
  \num{3} &  \num{73} & \num{2} &  NAYY $4\times120$ SE & \num{109.5} \\
  \num{6} &  \num{43} & \num{1} &  NAYY $4\times50 $ SE &  \num{81.5} \\
  \num{6} &  \num{68} & \num{2} &  NAYY $4\times120$ SE & \num{109.5} \\
 \num{28} &  \num{29} & \num{1} &  NAYY $4\times50 $ SE &  \num{81.5} \\
 \num{29} &  \num{30} & \num{1} &  NAYY $4\times50 $ SE &  \num{81.5} \\
 \num{50} &  \num{57} & \num{1} &  NAYY $4\times120$ SE &  \num{87.9} \\
 \num{50} &  \num{61} & \num{1} &  NAYY $4\times120$ SE &  \num{87.9} \\
 \num{56} &  \num{57} & \num{1} &  NAYY $4\times120$ SE &  \num{87.9} \\
 \num{63} &  \num{70} & \num{2} &  NAYY $4\times120$ SE & \num{109.5} \\
 \num{63} &  \num{69} & \num{2} &  NAYY $4\times120$ SE & \num{109.5} \\
 \num{65} &  \num{70} & \num{2} &  NAYY $4\times120$ SE & \num{109.5} \\
 \num{65} & \num{106} & \num{1} &  NAYY $4\times120$ SE &  \num{87.9} \\
 \num{68} & \num{100} & \num{2} &  NAYY $4\times120$ SE & \num{109.5} \\
 \num{69} & \num{100} & \num{2} &  NAYY $4\times120$ SE & \num{109.5} \\
\num{104} & \num{105} & \num{2} &  NAYY $4\times120$ SE & \num{109.5} \\
\bottomrule
\end{tabular}
\end{table}

\begin{figure}
    \centering
    \includegraphics[width=.95\linewidth]{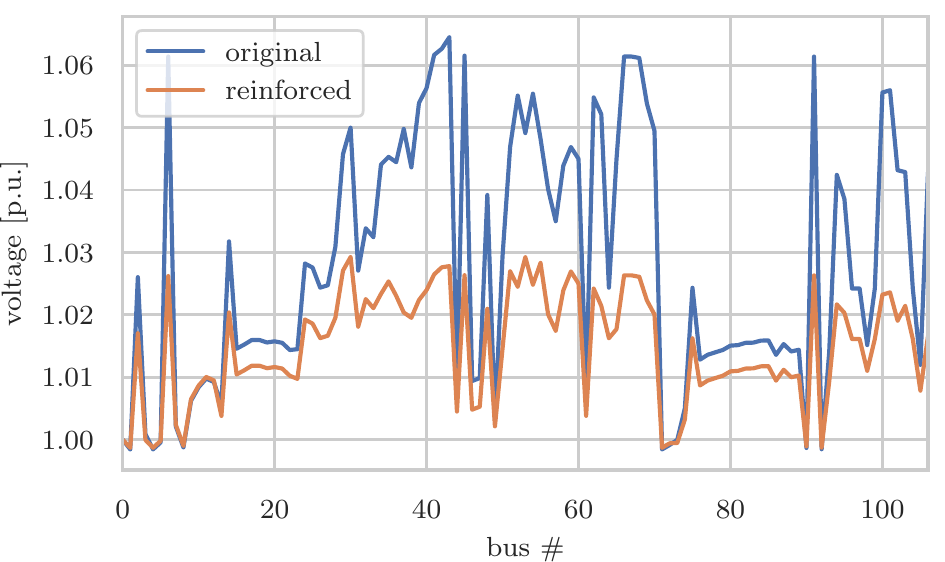}
    \vspace*{-3mm}
    \caption{Voltage comparison of the original and the reinforced grid. In this case a voltage limit of 1.03\,p.u. was used. }
    \label{fig:voltage_comp_re}
\end{figure}

The grid reinforcement algorithm has shown to be computationally effective and at the same time sufficient to fulfill the task of calculating a cost-effective grid reinforcement.
As it is a heuristic algorithm, the global optimality of the solution cannot be guaranteed. 
For radial distribution grids although a high-quality solution can be assumed. In Fig.\,\ref{fig:voltage_comp_re} a comparison of the bus voltages of the original grid and the reinforced one is displayed. It can be seen that the algorithm added additional elements to reinforce the grid until the voltage limits are met at all nodes.
The resulting reinforcement is shown in Fig.\,\ref{fig:re_grid}. The red edges of the graph indicate the reinforced line segments. Only the line segments with voltage problems were reinforced, in this case, three out of four branches. 

The selection of the line types depends on the amount of reinforcement needed to meet the voltage constraints. 
In this example, the algorithm chose to use 18 times a cable of the type NAYY $4\times120$ SE and three times the smaller NAYY $4\times50$ SE (see Table\,\ref{tab:line_re}). As the transformer loading was still within the limits it was not replaced or reinforced.

\subsection{Battery Placement}
\label{sec:batt_placement}

\begin{figure}
    \centering
    \includegraphics[width=.85\linewidth]{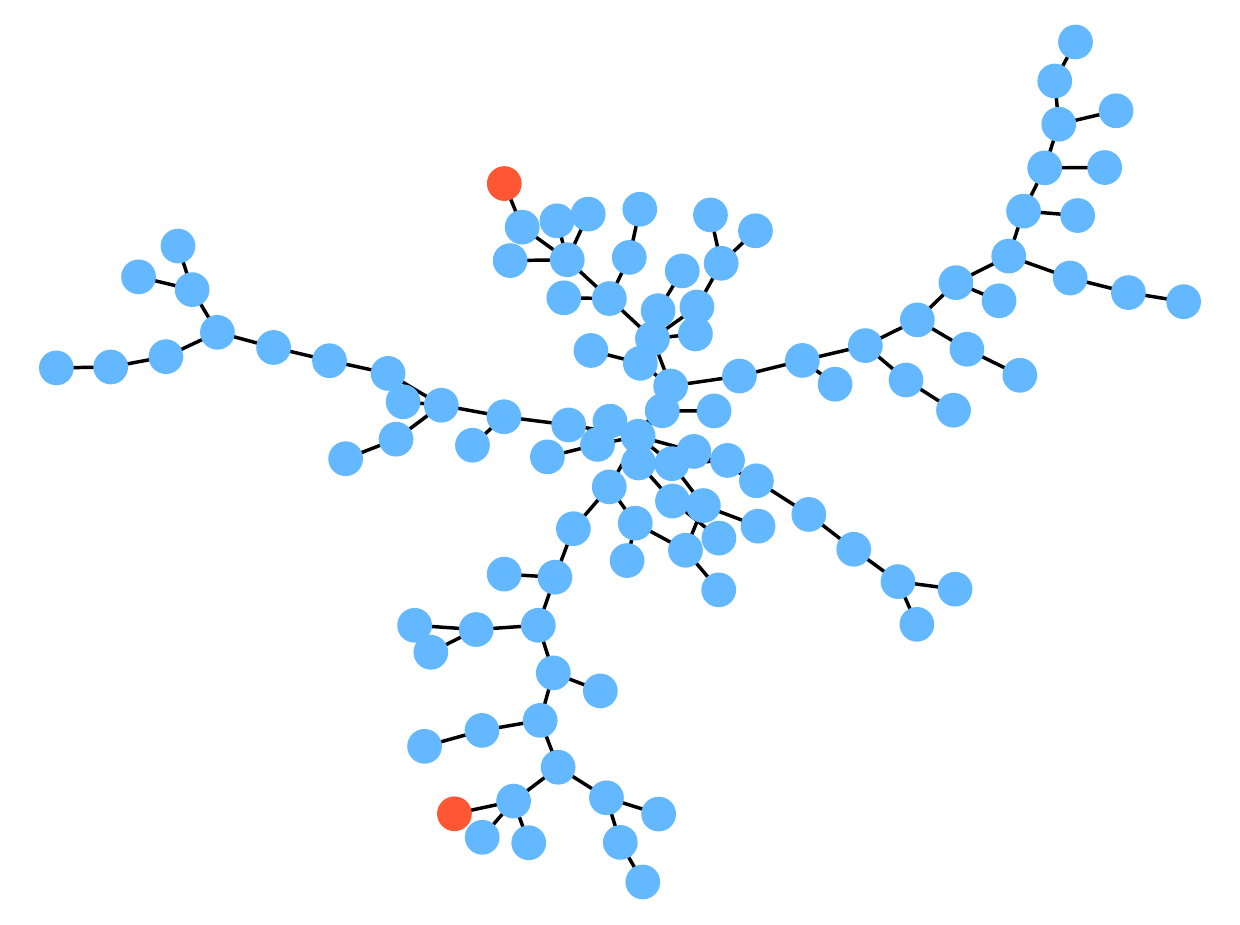}
    \vspace*{-3mm}
    \caption{Battery placement for a PV penetration of \SI{50}{\percent} $P_\text{PV, max}$ and a maximum voltage deviation of \SI{3}{\percent}.}
    \label{fig:batt_placement_05}
\end{figure}

\begin{figure}
    \centering
    \includegraphics[width=.85\linewidth]{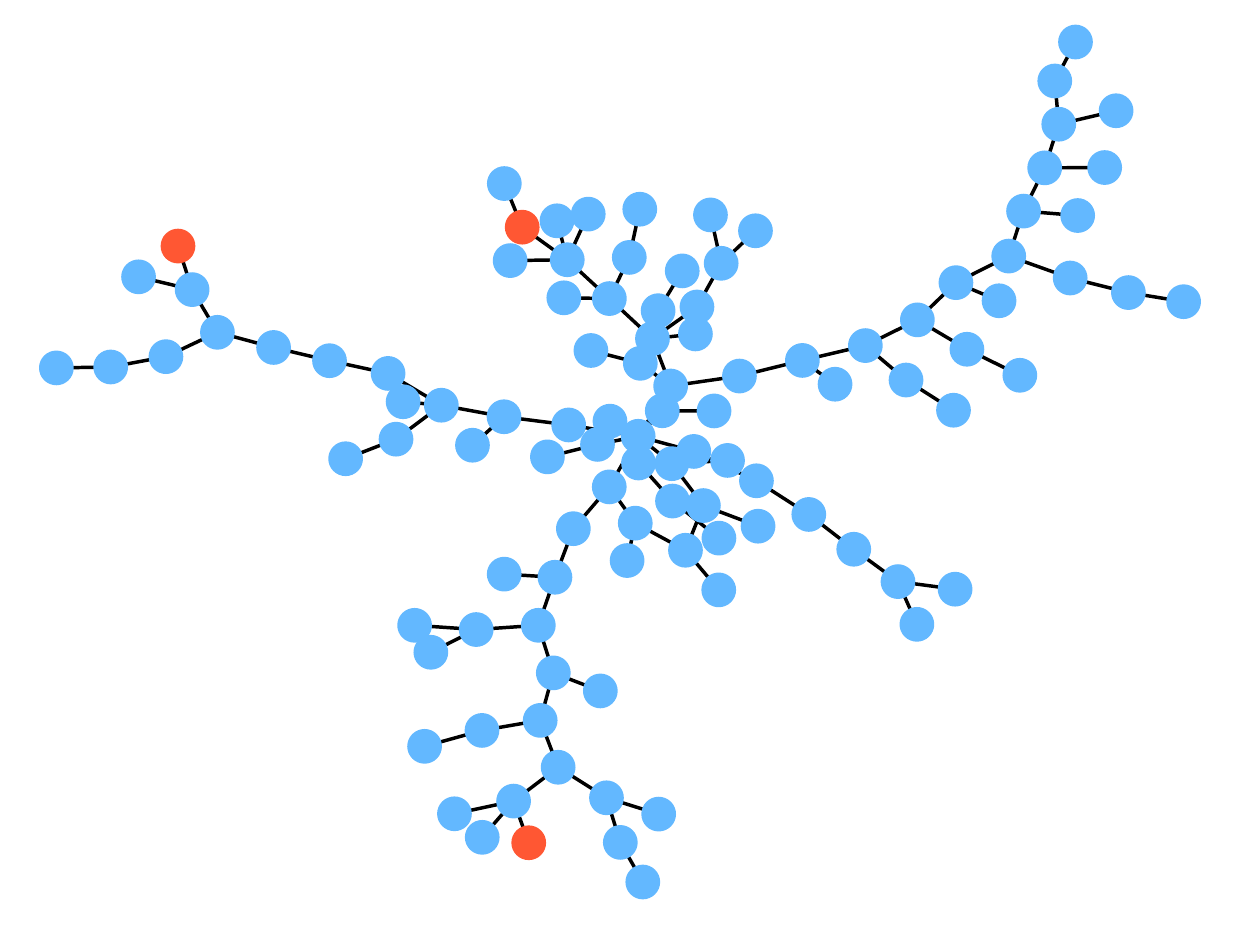}
    \vspace*{-3mm}
    \caption{Battery placement for a PV penetration of \SI{80}{\percent} $P_\text{PV, max}$ and a maximum voltage deviation of \SI{3}{\percent}.}
    \label{fig:batt_placement_03}
\end{figure}

The results of the battery placement are summed up in Table\,\ref{tab:size_loc} for the different scenarios. For the scenario of a \SI{50}{\percent} penetration and a maximum voltage deviation of \SI{5}{\percent} relative to $V_\text{nom}$ no battery placement is needed. If the maximum voltage deviation is decreased to \SI{3}{\percent}, 2 batteries have to be placed.
For the scenario with \SI{80}{\percent} PV penetration a maximum voltage deviation of \SI{5}{\percent} relative to $V_\text{nom}$,
the locations of the batteries stay almost the same as before, but the sizes increase.
If the maximum voltage deviation is decreased to \SI{3}{\percent} in this case, a third battery has to be added. The results are additionally shown in Fig.\,\ref{fig:batt_placement_05} and Fig.\,\ref{fig:batt_placement_03}. Here the placements for the \SI{80}{\percent} PV penetration and both voltage limitations are shown in a graph. 

\begin{table}[!h]
	\centering
 	\caption{Location and Size of the Battery}

\setlength{\tabcolsep}{4pt}
\renewcommand\cellgape{\Gape[4pt]}

\begin{tabular}{ccrrrrrr}
 \toprule
PV pen. & $\Delta\vect{V}_\text{max}$ & \mc{batt. 1} & \mc{batt. 2} & \mc{batt. 3}    \\
\cmidrule(rl){1-1}\cmidrule(rl){2-2}\cmidrule(rl){3-4}\cmidrule(rl){5-6}\cmidrule(rl){7-8}
[$P_\text{PV, max}$] & [$V_\text{nom}$] & C\,[kWh] & bus \#& C\,[kWh] & bus \# & C\,[kWh] & bus \# \\  
\midrule
\mr{50\,\si{\percent}}  & 5\,\si{\percent} & -  & -     & -     & -     & -     & - \\
                        & 3\,\si{\percent} & 57 & 30    & 120    & 42   & -     & -  \\[5pt]

\mr{80\,\si{\percent}} & 5\,\si{\percent} & 68  & 30     & 149   & 43   & -     & - \\
                       & 3\,\si{\percent} & 497 & 29    & 426   & 45    & 116   & 59 \\                        
 \bottomrule
\end{tabular}
	\label{tab:size_loc}
\end{table}

In all cases the placement of the battery stays almost the same but varies just slightly by one or two buses. This indicates that the solution surface is relatively flat with respect to the battery position in the grid.

\subsection{Comparison}

In this section, the results of the battery placement and the grid reinforcement algorithm will be compared.
In contrast to the previous section (\ref{sec:batt_placement}) some constraint on the number of batteries to place are added to investigate the sensitivity on the number of batteries installed.
Therefore two additional scenarios are introduced: one where the number of batteries is fixed to 5 and one with 10 batteries.

\begin{table}
\caption{Investment Cost Comparison}\label{tab:invest_comp}
\centering

\setlength{\tabcolsep}{6pt}
\begin{tabular}{cccccrc}
\toprule
PV pen. &
$\Delta \vect{V}_\text{max}$ &
grid reinf. & 
5 batt. & 
10 batt. & 
\mc{{unconstr.}} \\
\cmidrule(rl){1-1}\cmidrule(rl){2-2}\cmidrule(rl){3-3}\cmidrule(rl){4-4}\cmidrule(rl){5-5}\cmidrule(rl){6-7}
[$P_\text{PV, max}$] & [$V_\text{nom}$] & [k\euro] & [k\euro] & [k\euro] & [k\euro] & [n batt.]  \\
\midrule
\mr{50\,\%}       & 3\%               & 710         & 113     & 163      & 83        & 2        \\
          & 5\%               &  -           & -       & -        & -          & -        \\[5pt]
\mr{80\,\%}       & 3\%               & 1679        & 290     & 340      & 287       & 3        \\
          & 5\%               & 488         & 104     & 154      & 74        & 2      \\ 

\bottomrule

\end{tabular}

\end{table}

The results regarding the absolute investment costs are shown in Table\,\ref{tab:invest_comp}. The cheapest solution for the given scenarios is the installation of batteries to maintain the grid voltage stability. The costs increase with additional batteries, although the individual size of the batteries decreases. The most expensive option regarding the investment costs is grid reinforcement.

A more interesting comparison is the yearly costs. In this case, the maintenance costs for the battery as well as the grid are neglected, as they vary greatly. As mentioned before, the lifetime for the battery is assumed to be \SI{10}{\year} and the lifetime for the cables \SI{40}{\year}.
The results of this comparison are shown in Table\,\ref{tab:yearly_cost_comp}. The result is similar to the investment costs, although the difference between the battery placement and the grid reinforcement is less pronounced.

\begin{table}
\caption{Yearly Cost Comparison (assumed lifetime: battery=10\,yr, cable=40\,yr)}\label{tab:yearly_cost_comp}
\centering
\setlength{\tabcolsep}{6pt}
\begin{tabular}{cccccrc}
\toprule
PV pen. &
$\Delta \vect{V}_\text{max}$ &
grid reinf. & 
5 batt. & 
10 batt. & 
\mc{{unconstr.}} \\
\cmidrule(rl){1-1}\cmidrule(rl){2-2}\cmidrule(rl){3-3}\cmidrule(rl){4-4}\cmidrule(rl){5-5}\cmidrule(rl){6-7}
[$P_\text{PV, max}$] & [$V_\text{nom}$] & [k\euro] & [k\euro] & [k\euro] & [k\euro] & [n batt.]  \\
\midrule
\mr{50\%}       & 3\%               & 18          & 11      & 16       & 8         & 2        \\
          & 5\%               & 0           & 0       & 0        & 0         & 0        \\[5pt]
\mr{80\%}       & 3\%               & 42          & 29      & 34       & 29        & 3        \\
          & 5\%               & 12          & 10      & 15       & 7         & 2        \\ 

\bottomrule

\end{tabular}

\end{table}

\section{Conclusion and Outlook}
In this paper, we presented two algorithms: one for the automated placement and sizing of battery storage systems within a distribution grid and another algorithm for automated grid reinforcements. Both algorithms have shown to work and fulfill their task and avoid grid overloading for the suggested implementations. These algorithms can now be used to evaluate the possible solutions for distribution grids with congestion problems.

The results presented in this paper are only valid for the given scenario. 
However, the algorithm is capable of incorporating additional congestion management options such as the curtailment of renewable generation, load shifting, etc. 
These possibilities will be elaborated in a following paper.

\renewcommand*{\UrlFont}{\rmfamily}
\balance
\printbibliography

\end{document}